\documentclass[conference, a4paper]{IEEEtran}
\IEEEoverridecommandlockouts
\usepackage{cite}
\usepackage[english]{babel}
\usepackage{framed}
\usepackage[normalem]{ulem}
\usepackage{amsthm}
\usepackage{amssymb}
\usepackage[scr=dutchcal,cal=euler]{mathalfa}
\usepackage{enumerate}
\usepackage[utf8]{inputenc}
\usepackage{hyperref}
\hypersetup{
   colorlinks=true,
   linkcolor=blue,
   citecolor=blue,
   filecolor=magenta,      
   urlcolor=cyan,
}
\usepackage{amsmath,amssymb,amsfonts}
\usepackage{algorithm,algorithmic}
\usepackage{graphicx}
\usepackage{mathtools}
\usepackage{gensymb}
\usepackage{xcolor}
\usepackage{subcaption}
\usepackage{tabularx,booktabs}
\usepackage{multirow,multicol}
\usepackage{caption}
\def\BibTeX{{\rm B\kern-.05em{\sc i\kern-.025em b}\kern-.08em
    T\kern-.1667em\lower.7ex\hbox{E}\kern-.125emX}}
\begin{document}

\title{Estimation and Fault Detection on Hydraulic System with Adaptive-Scaling Kalman and Consensus Filtering\\
}

\author{Moh Kamalul Wafi$^{1}$ 
\thanks{$^{1}$Moh Kamalul Wafi is with Laboratory of Embedded \& Cyber-Pysical Systems, Department of Engineering Physics,
        Institut Teknologi Sepuluh Nopember, 60111, Indonesia,
        {\tt\small kamalul.wafi at its.ac.id}}
}

\maketitle
\begin{abstract}
The area of fault detection is becoming more interesting since there have been many unique designs to detect or even compensate the faults, either from sensor or actuator. This paper applies the hydraulic system with interconnected tanks by implementing a leakage on one of the three tanks. The mathematical model along with the details of stability properties are highly discussed in this paper by imposing the Lyapunov and boundedness stability. The theory of fault detection with certain threshold after the occurrence of the fault corresponding to the state estimation error is mathematically presented ended by simulation. Moreover, the system compares the effectiveness of the proposed observer using Luenberger observer, adaptive-scaling Kalman and consensus filtering. The results for some different initial condition guarantee the detection of the fault for some time $t_d > t_f$
\end{abstract}


\section{Introduction}
Hydraulic system is becoming more complex leading to various application. The idea behind this paper is presented in \cite{R1} with conventional estimation method. The importance of identify the model with certain control methods, fault scenario and frequency-domain identification is conducted in \cite{R2}, \cite{R3} and \cite{R4} in turn. Those designs somehow lead to beneficial approach to the real system, such as the dynamic in underwater characteristic \cite{R5}. More specifically, this fault detection and identification, due to leakage, in hydraulic system is highly studied. The analyses using adaptive robust approach, non-linear learning approach, offline feedback and adaptive control and unknown input of observer are done in \cite{R6}, \cite{R7}, \cite{R7}, and \cite{R8} respectively. Indeed, the changes are implementable and depended on the true system being observed, including the larger dynamic system of hydro-power plant as presented in \cite{R9}. However, those ideas are always initiated by the stability analysis before going through complex scenarios. 

One of the most famed stability properties is Lyapunov which is described by \cite{R10} along with certain networked system by \cite{R11}. After those formulas are fulfilled, the following is to design adaptive-Kalman which could be massively applied in divergent system, such as in cooperative localizaton \cite{R12}, motion \cite{R13}, along with MR-thermometry guided HIFU \cite{R14}. However, this paper not only discusses the adaptive-Kalman, rather using the more dynamic adaptive-scaling as initially proposed by \cite{R15} which is also used by \cite{R16} with absolutely the same scenario yet more specific application. Beyond that, the evaluation of adaptive Kalman being conducted in \cite{R17} is considered for some cases. By contrast, the consensus filtering proposed by \cite{R18} which is also presented in \cite{R19} is applied to observe the effectiveness of the adaptive-scaling. The paper is stated from the mathematical model followed by the estimation algorithms ended by some numerical simulation. 

\section{Mathematical Model}
\begin{figure}[!h]
    \centering
    \includegraphics[width=9cm]{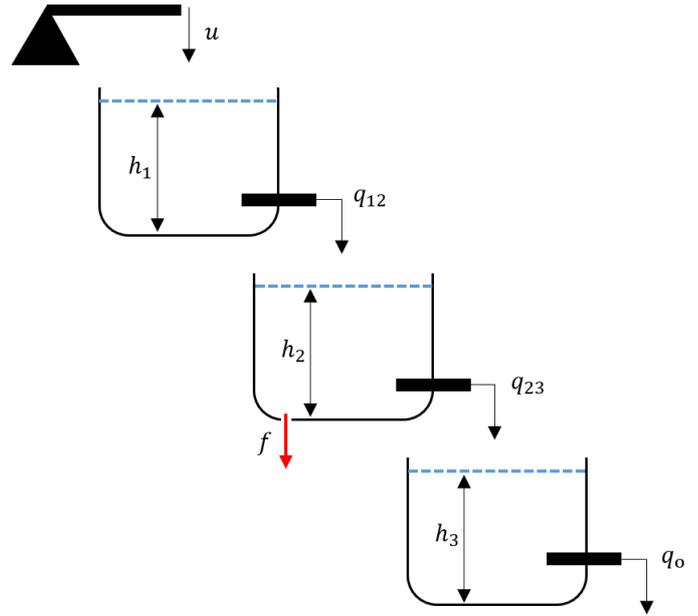}
    \caption{Hydraulic system scheme}
    \label{Fig1}
\end{figure}

The proposed design is the hydraulic system consisting of three tanks as illustrated in Fig. (\ref{Fig1}).
The values of $h_1$, $h_2$, and $h_3$ represent the fluid level in each tank while their volume of fluid in certain tank are denoted by
\begin{align}
    \mathcal{V}_{\tau} = h_{\tau}\psi_{\tau} \longrightarrow \tau = 1,2,3 \label{Eq. 1}
\end{align}
respectively. Moreover, for the sake of the linear relationships, those tanks are supposed to be "infinite" in terms of height so that the output flows have the mathematical distribution as follows,
\begin{align}
    q_{12} = \delta_1 h_1, \quad q_{23} = \delta_2 h_2, \quad q_{o} = \delta_3 h_3 \label{Eq. 2}
\end{align}
where the scalars of $\delta_{\tau}$ with $\tau = 1,2,3$ are set to be positive-known. The $u$ denotes the input flow while the sensor of level is designed in the third tank meaning that the system output is obtained from the the third tank of the fluid level written as $y = h_3$. The fault in the form of leakage $f$ is possibly situated in the second tank and such a fault as an output flow is assumed to happen at several unknown time-instant $t_f$, comprising
\begin{align}
    f(t) = \begin{cases}
    \bar{\delta}h_2, & t\geq t_f \\
    \\
    0, & \textrm{otherwise} \end{cases} \label{Eq. 3}
\end{align}
where $\bar{\delta}$ as the fault magnitude of the leakage is an positive-unknown constant. To solve this unhealthy system, it can be expressed by mathematical analysis and simulation to prove the hand-made calculation. The standard form of the state equation of this hydraulic system is written by
\begin{subequations}
\begin{align}
    \dot{x} &= Ax + Bu + \mathcal{F}f  \label{Eq. 4a}\\
    y &= Cx \label{Eq. 4b}
\end{align}
\end{subequations}
where $x, u, f$ represent the vector states, input, and fault of the system while the dimension of the matrices are $A \in \mathbb{R}^{n\times n}$, $B \in \mathbb{R}^{n\times p}$, $\mathcal{F} \in \mathbb{R}^{n\times p}$ and $C \in \mathbb{R}^{p\times n}$. The three states of $x_\tau$ portray the volume parameters of $\mathcal{V}_{\tau}$ for tank $(\tau)$, therefore
\begin{subequations}
\begin{gather}
    x_\tau \coloneqq \mathcal{V}_{\tau} = h_\tau\psi_\tau \longrightarrow \tau = 1,2,3 \label{Eq. 5a}\\
    \dot{x}_\tau\coloneqq\dot{\mathcal{V}}_{\tau} \label{Eq. 5b}
\end{gather}
\end{subequations}
so that from those concepts, the whole matrices containing the parameters is given by
\begin{align}
    \dot{x} &= \begin{bmatrix*}[c]
        -\dfrac{\delta_1}{\psi_1} & 0 & 0 \\[7pt]
        \dfrac{\delta_1}{\psi_1} & -\dfrac{\delta_2}{\psi_2} & 0 \\[7pt]
        0 & \dfrac{\delta_2}{\psi_2} & -\dfrac{\delta_3}{\psi_3}
        \end{bmatrix*} \begin{bmatrix}
        x_1 \\
        x_2 \\
        x_3 \end{bmatrix} + \begin{bmatrix}
        1 \\
        0 \\
        0 \end{bmatrix}u + \begin{bmatrix}
        0 \\
        -1 \\
        0 \end{bmatrix} f \label{Eq. 6}\\
        y &= \begin{bmatrix}
        0 & 0 & \dfrac{1}{\psi_3} \end{bmatrix} \begin{bmatrix}
        x_1 \\
        x_2 \\
        x_3 \end{bmatrix} \label{Eq. 7}
\end{align}

\subsection{Transfer Function}
The two models of transfer function of the system are then initiated with introducing a new matrix of $\bar{A}$, therefore
\begin{align}
    \bar{A} &= A + \mathcal{F}\textbf{I} \label{Eq. 8}
\end{align}
with $\bar{A} \in \mathbb{R}^{n\times n}$ and $\textbf{I} \in \mathbb{R}^{p\times n}$ is \textbf{ones} matrix. Keep in mind that the transfer function is written from two polynomials in the form of Laplace so that if
\begin{subequations}
\begin{align}
    \eta(s) &= s^m + \eta_1 s^{m-1} + \eta_2 s^{m-2} + \cdots + \eta_{m-1} s + \eta_m \label{Eq. 9a}\\
    \zeta(s) &= \zeta_1 s^{m-1} + \zeta_2 s^{m-2} + \cdots + \zeta_{m-1} s + \zeta_m \label{Eq. 9b}
\end{align}
\end{subequations}
and the roots of $\eta$ and $\zeta$ are the zeros and poles in turn of the system, then the transfer function $\Phi$ from those two relationships can be denoted as
\begin{align}
    \Phi(s) = \frac{\zeta(s)}{\eta(s)} \label{Eq. 10}
\end{align}
By giving the input $\eta(t)$ equals to $e^{st}$ where $s \neq \lambda_{\tau} (\bar{A}) \rightarrow \tau = 1, \cdots, m$ with $\lambda_{\tau}(\bar{A})$ describes the $\tau_{th}$ eigenvalue of $\bar{A}$. The states of the system, by choosing the initial state of $x(0)$ as $\left(sI - \bar{A}\right)^{-1}B$ is then written as follows,
\begin{align}
    x(t) &= e^{\bar{A}t}x(0) + e^{\bar{A}t} \int_0^{t} e^{\left(sI - \bar{A}\right)\tau}B \, d\tau \nonumber\\
    &= \left(sI - \bar{A}\right)^{-1}B e^{st} \label{Eq. 11}
\end{align}
and by substituting $x(t)$ into the output $\zeta(t)$ which is also $y(t)$, it can be said that
\begin{align}
    \zeta(t) \coloneqq y(t) &= Cx(t) \nonumber\\
    &= \Phi(s)\eta(t) \longrightarrow \Phi(s) = C\left(sI  -\bar{A}\right)^{-1}B \label{Eq. 12}
\end{align}
and thanks to (\ref{Eq. 12}), the transfer function of the "healthy" system is denoted as follows
\begin{align}
    \Phi_{\mathcal{H}}(s) = \frac{\left(\dfrac{\delta_1}{\psi_1}\dfrac{\delta_2}{\psi_2}\right)\dfrac{1}{\psi_3}}{\left(s + \dfrac{\delta_1}{\psi_1}\right)\left(s + \dfrac{\delta_2}{\psi_2}\right)\left(s + \dfrac{\delta_3}{\psi_3}\right)} \label{Eq. 13}
\end{align}
and for the "unhealthy" system, it is given by considering the appearance of $\bar{\delta}$, such that,
\begin{align}
    \Phi_{\upsilon}(s) = \frac{\left(\dfrac{\delta_1}{\psi_1}\dfrac{\delta_2}{\psi_2}\right)\dfrac{1}{\psi_3}}{\left(s + \dfrac{\delta_1}{\psi_1}\right)\left(s + \dfrac{\delta_2 + \bar{\delta}}{\psi_2}\right)\left(s + \dfrac{\delta_3}{\psi_3}\right)}\label{Eq. 14}
\end{align}
Bear in mind that the important of invertible matrix from $\left(sI - \bar{A}\right)^{-1}$ is crucial with $s \neq \lambda_{\tau}(\bar{A})$, otherwise the singular matrix occurs.

\subsection{Stability Properties}
Stability properties are the further analysis being conducted. The two most-famed stability have been well-defined by Lyapunov stability and bounded-input \& bounded-output stability, representing the internal and external stability of such system respectively. The relationship between these two is compact in LTI system whereas for the time-varying system, they should be put into the canonical form. Recalling the (\ref{Eq. 4a}) and (\ref{Eq. 4b}), they are denoted as "uniformly asymptotically stable" if the $\varphi(t, \epsilon)$ related to those equations guarantees the following inequality,
\begin{align}
    \left\Vert\varphi(t, \epsilon)\right\Vert \leq Q_1e^{Q_2(t - \epsilon)} \label{Eq. 15}
\end{align}
with positive constants of those $Q$. The bounded-input \& bounded output is achieved if $\left\Vert u(t)\right\Vert \leq Q_3$, $\forall (t,u)$ and $Q_3$-positive satisfying $\left\Vert x(t)\right\Vert < Q_4\left(Q_3, \left\Vert x(t_0)\right\Vert\right)$ for $t \geq t_0$. This equals to the condition,
\begin{align}
    \int_{-\infty}^{t} \left\Vert\varphi(t, \gamma)B(\gamma)\right\Vert\, d\gamma \leq Q_5 \label{Eq. 16}
\end{align}
where $\left\Vert\varphi(t, t_0)\right\Vert \leq Q_6$, $\forall t$ and positive constants of $Q_5, Q_6$. This could create the coincidence of both stability unless further assumptions are situated in the matrices of $A, B$. It is should be guaranteed that the $a_i, (i = 1, \ldots, n)$ in the companion matrix of $(A,B)$ defined as $(\mathcal{F}, \mathcal{G})$ are bounded,
\begin{align*}
    \mathcal{F} = \begin{bmatrix*}[c]
    0 & 1 & 0 & \star & \star \\
    0 & 0 & 1 & \star & \star \\
    \star & \star & \star & \star & \star \\
    \star & \star & \star & 0 & 1 \\
    -\alpha_1 & -\alpha_2 & \star & -\alpha_{n-1} & -\alpha_n \end{bmatrix*} \quad \mathcal{G} = \begin{bmatrix}
    0 \\
    \star \\
    \star \\
    \star \\
    1 \end{bmatrix}
\end{align*}
By holding this boundedness, the feedback law of $u = -Kx$ exists leading to the connection of the eigenvalues $\lambda_i$ and $\mu_i$ of $\mathcal{F}$ and $\mathcal{F}-\mathcal{G}K$ in turn by,
\begin{align}
    \mu_i = \lambda_i + \alpha \longrightarrow i = 1, 2, \ldots, n \label{Eq. 17}
\end{align}
where $\alpha$ is the arbitrary and by introducing the $\mathcal{T}_T$ matrix  as
\begin{align}
    \mathcal{T}_T = \begin{bmatrix*}[c]
    1 & 0 & \star & \star & \star \\
    -\alpha & 1 & 0 & \star & \star \\
    \alpha^2 & -2\alpha & 1 & 0 & \star \\
    -\alpha^3 & 3\alpha^2 & -3\alpha & 1 & 0 \\
    \star & \star & \star & \star & 1 \end{bmatrix*} \label{Eq. 18}
\end{align}
holds $\mathcal{T}_T(\mathcal{F} - \mathcal{G}K)\mathcal{T}_T^{-1} = \alpha I + \mathcal{F}$. Moreover, suppose that it is allowed to maintain the $x(t_0)$ with $\left\Vert x(t_0)\right\Vert \leq 1$ which makes $\left\Vert u\right\Vert < K_3$ leading to $\left\Vert x\right\Vert < K_4(K_3)$ for which $K_4 > 1$. Suppose that $\alpha$ is maintained small such that $\left\Vert x\right\Vert < K_4$ implies $\left\Vert Kx\right\Vert < K_3$, $\forall t$ therefore the closed-loop design formed as $\dot{x} = \left(\mathcal{F} - \mathcal{G}K\right)x$ comprises Lyapunov stability. Since $\left\Vert u\right\Vert < K_3$ and due to $\mathcal{T}_T$, it constitutes $\left\Vert x(\mathcal{T})\right\Vert = K_4$ and this satisfies $\left\Vert x(t)\right\Vert < K_4$ ranging from $[t_0, \mathcal{T}]$, such that,
\begin{align}
    \dot{x} = \left(\alpha I + \mathcal{F}\right)x \label{Eq. 19}
\end{align}
for which this follows that the system is "exponentially asymptotically stable". Moving to other properties, in a trivial way, two matrices are categorized as controllable if and only if the rank $(\rho)$ of the controllability matrix $\mathcal{Q}_C$ is exactly on par with that of in the $A$ matrix. Controllability matrix is defined as the combinations of both $A$ and $B$ matrix with $A$ as $n$ by $n$ matrix, such that:
\begin{align}
	\mathcal{Q}_C = \left.\begin{cases}\begin{bmatrix}
	B & AB & \cdots & A^{n-1}B
	\end{bmatrix} \\
	\textrm{or} \\
	\begin{bmatrix}
	B & \bar{A}B & \cdots & \bar{A}^{n-1}B
	\end{bmatrix}\end{cases}\right\} \rightarrow \rho_C = n \label{Eq 20}
\end{align}
where $\mathcal{Q}_C(\mathcal{H})$ and $\mathcal{Q}_C(\upsilon)$ are denoted as the controllability matrices of healthy and faulty system in turn, therefore,
\begin{subequations}
\begin{align}
    \mathcal{Q}_C(\upsilon) = & \begin{bmatrix*}[c]
	1 & -\dfrac{\delta_1}{\psi_1} & \left(\dfrac{\delta_1}{\psi_1}\right)^2 \\
	0 & \dfrac{\delta_1}{\psi_1} & -\left(\dfrac{\delta_1}{\psi_1}\right)^2 - \dfrac{(\delta_1\delta_2 + \delta_1\bar{\delta})}{\psi_1\psi_2} \\
	0 & 0 & \dfrac{\delta_1\delta_2}{\psi_1\psi_2} \end{bmatrix*} \label{Eq 21a}\\
	\mathcal{Q}_C(\mathcal{H}) = & \begin{bmatrix*}[c]
	1 & -\dfrac{\delta_1}{\psi_1} & \left(\dfrac{\delta_1}{\psi_1}\right)^2 \\
	0 & \dfrac{\delta_1}{\psi_1} & -\left(\dfrac{\delta_1}{\psi_1}\right)^2 - \dfrac{\delta_1\delta_2}{\psi_1\psi_2} \\
	0 & 0 & \dfrac{\delta_1\delta_2}{\psi_1\psi_2} \end{bmatrix*} \label{Eq 21b}
\end{align}
\end{subequations}
Observability $\mathcal{Q}_O$ performs on par with controllability regarding the similarity rank of to the system of either $A$ of $\bar{A}$ matrix. Nevertheless, instead of applying $B$, this uses $C$ arrangement and also the observability matrix is arranged in the one column, such that: (suppose $A$ is $n$ by $n$ matrix)
\begin{equation}
	\mathcal{Q}_O = \begin{bmatrix}
	C \\
    CA \\
    \vdots \\
    CA^{n-1}
	\end{bmatrix} \textrm{ or } \begin{bmatrix}
	C \\
    C\bar{A} \\
    \vdots \\
    C\bar{A}^{n-1}
	\end{bmatrix} \longrightarrow \rho_O = n
    \label{Eq 22}
\end{equation}
where the obsevability matrices of healthy $\mathcal{Q}_O(\mathcal{H})$ and faulty $\mathcal{Q}_O(\upsilon)$ system is explored in the following, therefore,
\begin{subequations}
\begin{align}
    \mathcal{Q}_O(\upsilon) = & \begin{bmatrix*}[c]
	0 & 0 & \dfrac{1}{\psi_3} \\[7pt]
    0 & \dfrac{\delta_2}{\psi_2\psi_3} & -\dfrac{\delta_3}{(\psi_3)^2} \\[7pt]
    \dfrac{\delta_1\delta_2}{\psi_1\psi_2\psi_3} & -\dfrac{(\delta_2)^2 + \delta_2\bar{\delta}}{(\psi_2)^2}-\dfrac{\delta_2\delta_3}{\psi_2\psi_3} & \dfrac{(\delta_3)^2}{(\psi_3)^3} \end{bmatrix*} \label{Eq 23a}\\
    \mathcal{Q}_O(\mathcal{H}) = & \begin{bmatrix*}[c]
	0 & 0 & \dfrac{1}{\psi_3} \\[7pt]
    0 & \dfrac{\delta_2}{\psi_2\psi_3} & -\dfrac{\delta_3}{(\psi_3)^2} \\[7pt]
    \dfrac{\delta_1\delta_2}{\psi_1\psi_2\psi_3} & -\dfrac{(\delta_2)^2} {(\psi_2)^2}-\dfrac{\delta_2\delta_3}{\psi_2\psi_3} & \dfrac{(\delta_3)^2}{(\psi_3)^3} \end{bmatrix*} \label{Eq 23b}
\end{align}
\end{subequations}

\subsection{Design of State Observer}
The state-space of Luenberger observer is given as follow with the aim to seek the best design for $\Psi$,
\begin{subequations}
\begin{align}
    \dot{\hat{x}} =\; & A\hat{x} + Bu + \Psi\left[y - \hat{y}\right]\label{Eq 24a}\\
    \hat{y} =\; & C\hat{x}\label{Eq 24b}
\end{align}
\end{subequations}
Hence, the observability canonical form from the healthy system (\ref{Eq. 13}) is given as,
\begin{subequations}
\begin{align}
    \dot{x} =\; & A_{o}x + B_{o}u \label{Eq 25a}\\
    y =\; & C_{o} x \label{Eq 25b}
\end{align}
\end{subequations}
where,
\begin{align*}
    A_{o} =\;& \begin{bmatrix*}[c]
    0 & 0 & -\dfrac{\delta_1\delta_2\delta_3}{\psi_1\psi_2\psi_3}\\[7pt]
    1 & 0 & -\left(\dfrac{\delta_1\delta_2}{\psi_1\psi_2} + \dfrac{\delta_1\delta_3}{\psi_1\psi_3} + \dfrac{\delta_2\delta_3}{\psi_2\psi_3}\right)\\[7pt]
    0 & 1 & -\left(\dfrac{\delta_1}{\psi_1} + \dfrac{\delta_2}{\psi_2} + \dfrac{\delta_3}{\psi_3}\right)\end{bmatrix*}\\
    B_{o} =\;& \begin{bmatrix}
    1 \\
    0 \\
    0 \end{bmatrix} \quad C_{o} = \begin{bmatrix}
    0 & 0 & 1\end{bmatrix} 
\end{align*}
and the observability matrix in the format of canonical form is defined as,
\begin{align}
    \bar{\mathcal{Q}}_O = \begin{bmatrix}
	C_o \\
    C_oA_o \\
    \vdots \\
    C_oA_o^{n-1}
	\end{bmatrix} \longrightarrow \rho_O = n \label{Eq 26}
\end{align}
Furthermore, the transformation matrix of $\Delta$ of the canonical form is obtained as,
\begin{align}
    \Delta = \mathcal{Q}_O(\mathcal{H})^{-1} \bar{\mathcal{Q}}_O \label{Eq 27}
\end{align}
The error dynamics of the system in the form of observer canonical form is denoted as $\Gamma$, for which
\begin{align}
    \Gamma = A_o - \Psi_o C_o \longrightarrow \Psi_o = \begin{bmatrix}
    l_1\\
    l_2\\
    l_3\end{bmatrix} \label{Eq 28}
\end{align}
and the characteristic polynomial of $\Gamma$ is $\det (sI - \Gamma)$. Proposing a new matrix of $\Gamma_d$ with the desired eigenvalues of $\lambda_i$ for $i =1, \ldots, n$ with $n$ equals to the number of full rank of $A$. From this, the polynomial of $\Gamma$ should be compared to that of $\Gamma_d$ with certain $\lambda_i$ and the value of $\Psi_o$ is then obtained. Lastly, the gain of the observer is given by,
\begin{align}
    \Psi = \Delta\Psi_o \label{Eq 29}
\end{align}
and the matrix showing the dynamic of the system is written by $\Gamma_d = A - \Psi C$.

\subsection{Adaptive-Scaling Kalman Filter}
The idea of the term "scaling" is proposed in \cite{R16} which is then further modified in \cite{R17}. The following is the prediction of a dynamic system in discrete time,
\begin{align}
	\hat{x}_{k|k-1} &= A_k \hat{x}_{k-1|k-1} + B_k \vartheta_k \label{Eq 30}\\
    \hat{y}_k &= \Theta_k \hat{x}_{k|k-1} + \chi_k \label{Eq 31}
\end{align}
where $\hat{y}_k \in \mathbb{R}^p$ as measurements whereas $\vartheta_k$ and $\chi_k$ are set as $\mathcal{W}\mathcal{G}\mathcal{N}$ such that $\textbf{E}(\vartheta_k \vartheta_l^\top) = Q_k\Omega$ and $\textbf{E}(\chi_k \chi_l^\top) = R_k\Omega$. $x_0$ is defined as the initial state with $x_0 = \mathcal{N}(\bar{x}_0, P_0)$. Given the set of measurements $Y_k = [y_0,y_1,\dots,y_k]$. From those, the covariance matrices of prediction in Kalman are then set as given
\begin{align}
    P_{k+1} &= AP_{k}A^\top + \phi_k B_k Q_k B_k^\top \label{Eq 32}\\
    \bar{P} &= \Theta P_{k+1}\Theta^\top + R_k \nonumber\\
    &= \Theta AP_{k}A^\top\Theta^\top + R_k + \phi_k \Theta B_k Q_k B_k^\top \Theta^\top \label{Eq 33}
\end{align}
where $\phi_k$ is what is called as scaling parameter as a modification of covariance matrices. Let $\nu_k = y_k - \hat{y}_{k|k-1}$ and, a-priori, the result of the $(\nu_k)^2$ being showed as $|\nu_k|^2 = \nu^\top\nu$ is supposed to be the diagonal elements of $\bar{P}$ consisting the scaling variable $\phi$. For the simplification, by introducing new variables $\alpha_\theta, \beta_\theta, \gamma_\theta$, the equation (\ref{Eq 33}) can be written as,
\begin{subequations}
\begin{align}
    \alpha_\theta &= \bar{P} \label{Eq 34a}\\
    \beta_\theta &= \Theta AP_{k}A^\top\Theta^\top + R_k \label{Eq 34b}\\
    \gamma_\theta &= \phi_k \Theta B_k Q_k B_k^\top \Theta^\top \label{Eq 34c}
\end{align}
\end{subequations}
such that,
\begin{align}
    \alpha_\theta = \beta_\theta + \gamma_\theta \label{Eq 35}
\end{align}
To obtain $\phi$, by proposing some new constants of $a,b,c$ with $a\geq0, b\geq0, c\geq0$ and $a+b+c = 1$, it is given by
\begin{align}
    \Upsilon_k = a\phi_0 + b\phi_{k-1} + c\left[\frac{\nu_k - (\beta_\theta)_k}{(\gamma_\theta)_k}\right] \label{Eq 36}
\end{align}
where $\phi_k = \max(\Upsilon_k,0)$ and $\phi_0 = 1$ at $k = 0$. if $|\nu_k|^2 = \alpha_\theta$,
\begin{align}
    \Upsilon_k &= a\phi_0 + (b + c)\phi_{k-1} \label{Eq 37}\\
    \phi_k &= \max(\Upsilon_k,0) \label{Eq 38}
\end{align}
The updates of the estimated states and covariance matrices are presented in the following,
\begin{align}
    K_k &= P_{k|k-1}\Theta^\top\bar{P}^{-1} \label{Eq 39}\\
    \hat{x}_{k|k} &= \hat{x}_{k|k-1} + K_k \nu_k \label{Eq 40}\\
    P_{k|k} &= (I - K_k\Theta)P_{k|k-1} \label{Eq 41}
\end{align}
Keep in mind that the this linear system could be switched into non-linear system as it is an adaptive-scaling Kalman filtering.

\subsection{Consensus Filtering}
From \cite{R18} and \cite{R19} and recalling (\ref{Eq 30}) and (\ref{Eq 31}) using the same scenarios of initial condition of covariance matrices $Q_k, R_k, x_0$ with the same dimensional definition of $Y_k$ and $\nu_k$, the state estimates could be written as,
\begin{align}
\begin{split}
    \hat{x}_k &= \textbf{E}(x_k|Y_k),\\
    P_k &= \sum({k+1|k}),
\end{split}
\qquad
\begin{split}
    \bar{x}_k &= \textbf{E}(x_k|Y_{k-1})\\
    \mathcal{P}_k &= \sum(k+1)
\end{split}\label{Eq 42}
\end{align}
where $P_k$ as initial state of error covariance. That $P_k$ and the inverse of covariance matrix $H_k$ with certain number of sensors communicated as distributed large-scale system are paired to generate matrix $\mathcal{P}_k$, such that
\begin{gather}
	H_k = \frac{1}{n}\sum_{i=1}^{n}\left(\Theta^\top_k R^{-1}_k \Theta_k\right)_i \label{Eq 43}\\
    \mathcal{P}_k^{-1} = (nP)^{-1} + H_k \label{Eq 44}
\end{gather}
whilst matrix of $K$ is affected by the matrix of $\mathcal{P}$ as given,
\begin{equation}
	K = \mathcal{P}_k \Theta^\top_k R^{-1}_k \label{Eq 45}
\end{equation}
The updates of the state estimation is formulated as follows:
\begin{align}
    \hat{x}_{k} &= \hat{x}_{k|k-1} + K \left(y_k - \Theta\hat{x}_{k|k-1}\right)\label{Eq 46}\\
	&= \hat{x}_{k|k-1} + \mathcal{P}_k (\Theta^\top_k R^{-1}_k y_k - \Theta^\top_k R^{-1}_k \Theta\hat{x}_{k|k-1})\label{Eq 47}
\end{align}
from the equation, it can be simplified by introducing $z_k$ which is also affected by some $n$ sensors, if any, and a new measurement is written as 
\begin{gather}
    z_k = \frac{1}{n}\sum_{i=1}^{n}\Theta^\top_k R^{-1}_k y_k\label{Eq 48}\\ 
	\hat{x}_m = \bar{x} + \mathcal{P}_k (z_k - H_k \bar{x}) \label{Eq 49}
\end{gather}
compared to the classic Kalman filter, the prediction of $\hat{x}$ and $P$ are becoming the update of this $\mu$KF such that:
\begin{align}
	P^+_k &= A \mathcal{P}_k A^\top + B Q_k B^\top \label{Eq 50}\\
    \bar{x}^+ &= A\hat{x}_m \label{Eq 51}
\end{align}
This algorithm has some pros in terms of simplifying a huge calculation in conventional-KF which is not feasible. While it is required some additional high-, low-, and band-pass filter in tackling consensus dynamic problems, for this paper, it is only halt in (\ref{Eq 51}). This method is comparable as another modified algorithm from the original KF and they are observed in terms of the capability in estimating the unhealthy system.

\begin{algorithm}
	\caption{Adaptive-Scaling and Consensus Filtering}
 	\begin{algorithmic}[0]
    \STATE \textbf{Initialization:}
    \STATE $x_0 \sim \mathcal{N}(\bar{x}_0, P_0)$; $q \sim \mathcal{N}(\bar{q}, \Psi_q)$;\\
    \STATE $R_k$, $Q_k$, $Y_k = [y_0\to y_k]$, $\alpha_\theta$, $\beta_\theta$, $\gamma_\theta$, $\phi_k$, $z_k$;\\
    \STATE $x_0 = \bar{x}_0$ and $P_0 = \Psi_x$
    \FOR{$i = 1$ to $\kappa$} 
    \FOR{$k = 1$ to $N$}
	\STATE \textbf{1. Adaptive-Scaling Kalman:}
    \begin{align*}
    	\hat{x}_{k|k-1} &= A_k \hat{x}_{k-1|k-1} + B_k \vartheta_k \\
        \hat{y}_k &= \Theta_k \hat{x}_{k|k-1} + \chi_k \\
        \alpha_\theta &= \beta_\theta + \gamma_\theta\\
    	\Upsilon_k &= \begin{cases}
    	a\phi_0 + b\phi_{k-1} + c\left[\frac{\nu_k - (\beta_\theta)_k}{(\gamma_\theta)_k}\right], \textrm{otherwise}\\
    	a\phi_0 + (b + c)\phi_{k-1}, \textrm{if } |\nu_k|^2 = \alpha_\theta
    	\end{cases}\\
    	\phi_k &= \max(\Upsilon_k,0)\\
    	K_k &= P_{k|k-1}\Theta^\top\bar{P}^{-1} \\
        \hat{x}_{k|k} &= \hat{x}_{k|k-1} + K_k \nu_k \\
        P_{k|k} &= (I - K_k\Theta)P_{k|k-1} 
    \intertext{\STATE \textbf{2. \textit{Consensus}-Filtering:}}
    	H_k &= \Theta^\top_k R^{-1}_k \Theta_k\\
    	\mathcal{P}_k &= (P^{-1} + H_k)^{-1}\\
    	z_k &= \Theta^\top_k R^{-1}_k y_k\\
        \hat{x}_m &= \bar{x}_0 + \mathcal{P}_k (\hat{z}_k - H_k \bar{x}_0)\\
        P_k^+ &= A\mathcal{P}_k A^\top + B Q_k B^\top\\
        \bar{x}^+ &= A\hat{x}_m
    \intertext{\STATE \textbf{Collecting Estimation Error:}}
    	e_1 &= |x_{k} - \hat{x}_k|\\
        e_2 &= |x_{k} - \bar{x}^+_k|
    \end{align*}
    \ENDFOR
    \ENDFOR
    \end{algorithmic} 
\end{algorithm}

\subsection{Fault Detection Scheme}
This subsection is initiated by computing the output residual which is then defined as,
\begin{align}
    \varepsilon(t) = Ce(t) \longrightarrow e(t) = \mathcal{L}^{-1}\left[(sI - \Gamma_d)^{-1}\right]e(0) \label{Eq 54}
\end{align}
where $e(t)$ is the state estimation error being described as $e(t) = x(t) - \hat{x}(t)$ and its dynamics is denoted as $\dot{e}(t) = \Gamma_d e(t)$. The fault detection scheme is supposedly detected at time $t_d > t_f$, satisfying $\varepsilon(t_d) > \bar{\varepsilon}(t_d)$ with
\begin{align}
    \bar{\varepsilon}(t) = C\mathcal{L}^{-1}\left[(sI - \Gamma_d)^{-1}\right]\hat{e} \longrightarrow \hat{e} = \begin{bmatrix}
    \hat{e}_1\\
    \hat{e}_2\\
    \hat{e}_3\end{bmatrix} \label{Eq 55}
\end{align}
where $\bar{\varepsilon}$ is the designed appropriate threshold with some initial ranges of $x_i(0)$ for $i = 1, \ldots, n$ allowing to detect the fault $f$ in a finite time $t_d > t_f$. Observing that $\varepsilon(t) \rightarrow 0$, $\forall t \rightarrow \infty$, it can be said that for $t \geq t_f, \exists t_d > t_f \coloneqq \varepsilon(t_d) > \bar{\varepsilon}(t_d)$.  

\section{Numerical Results}
The design of the simulation is shown with the following variables so that $\psi_1 = 2$, $\psi_2 = 1$, $\psi_3 = 2$ and $\delta_1 = 1$, $\delta_2 = 1.5$, $\delta_3 = 1$. From these, the state space system is,
\begin{align*}
    \dot{x} &= \overbrace{\begin{bmatrix*}[r]
        -0.5 &  0.0 & 0.0 \\
         0.5 & -1.5-\bar{\delta} & 0.0 \\
         0.0 &  1.5 & -0.5
        \end{bmatrix*}}^{\bar{A}}x + \begin{bmatrix}
        1 \\
        0 \\
        0 \end{bmatrix}u \\
        y &= \begin{bmatrix}
        0 & 0 & 0.5 \end{bmatrix} x
\end{align*}
and removing $\bar{\delta}$ to get the healthy matrix of system $A$ with the transfer function of the healthy system is defined as follows,
\begin{align*}
    \Phi_{\mathcal{H}}(s) = \frac{0.375}{s^3 + 2.5s^2 + 1.75s + 0.375}
\end{align*}
and for the unhealthy, it is given by,
\begin{align*}
    \Phi_{\upsilon}(s) = \frac{0.375}{s^3 + (2.5 + \bar{\delta})s^2 + (1.75 + \bar{\delta})s + 0.375 + 0.25\bar{\delta}}
\end{align*}
From the transfer functions created in (\ref{Eq. 13}) and (\ref{Eq. 14}), the stability properties can be examined. For the healthy system, since $\delta_i$ and $\psi_i$ are positive constants, the poles of the $\det (sI - A)$ are all negative so that the system is,
\begin{align*}
    \det (sI - A) = \left(s + 0.5\right)\left(s + 1.5\right)\left(s + 0.5\right)
\end{align*}
asymptotically stable, whereas for the faulty system with $\bar{\delta}$ as positive scalar with $\bar{\delta}>0$ constitutes also asymptotically stable, such that
\begin{align*}
    \det (sI - A) = \left(s + 0.5\right)\left(s + 1.5 + \bar{\delta}\right)\left(s + 0.5\right)
\end{align*}
The controllablity matrices of the two conditions are illustrated as in (\ref{Eq 21a}) and (\ref{Eq 21b}), such that
\begin{align*}
    \mathcal{Q}_C(\mathcal{H}) =& \begin{bmatrix*}[r]
    1 & -0.5 &  0.25 \\
    0 &  0.5 & -1.00 \\
    0 &  0.0 &  0.75 \end{bmatrix*} \\
    \mathcal{Q}_C(\upsilon) =& \begin{bmatrix*}[r]
    1 & -0.5 &  0.25 \\
    0 &  0.5 & -1-0.5\bar{\delta} \\
    0 &  0.0 &  0.75 \end{bmatrix*}
\end{align*}
where the ranks of both $\mathcal{Q}_C(\mathcal{H})$ and $\mathcal{Q}_C(\upsilon)$ are $n = 3$ with $\bar{\delta}>0$. Moreover, the observability matrices of both kind of systems are presented from (\ref{Eq 23a}) and (\ref{Eq 23b}) as follows, 
\begin{align*}
    \mathcal{Q}_O(\mathcal{H}) =& \begin{bmatrix*}[r]
    0.000 &  0.000 &  0.500 \\
    0.000 &  0.750 & -0.250 \\
    0.375 & -1.500 &  0.125 \end{bmatrix*} \\
    \mathcal{Q}_O(\upsilon) =& \begin{bmatrix*}[r]
    0.000 & 0.000 & 0.500 \\
    0.000 & 0.750 & -0.250 \\
    0.375 & -1.5-0.75\bar{\delta} & 0.125 \end{bmatrix*}
\end{align*}
with the full rank for either two matrices $\rho_O = 3$. Recalling to (\ref{Eq 25a}) and (\ref{Eq 25b}) with respect to estimation, the set matrices are
\begin{align*}
    A_o = \begin{bmatrix*}[l]
    0 & 0 & -0.375\\
    1 & 0 & -1.750\\
    0 & 1 & -2.500\end{bmatrix*}, \quad B_o = \begin{bmatrix}
    1 \\
    0 \\
    0 \end{bmatrix}, \quad C_o = \begin{bmatrix}
    0 & 0 & 1\end{bmatrix}
\end{align*}
such that the canonical form of its observability $\bar{\mathcal{Q}}_O$ along with the transformation matrix $\Delta$ is
\begin{align*}
    \bar{\mathcal{Q}}_O = \begin{bmatrix*}[r]
    0.0 &  0.0 &  1.0\\
    0.0 &  1.0 & -2.5\\
    1.0 & -2.5 &  4.5\end{bmatrix*}, \quad \Delta = \begin{bmatrix*}[r]
    \frac{8}{3} & -\frac{4}{3} & \frac{2}{3}\\[5pt]
    0 & \frac{4}{3} & -\frac{8}{3}\\[5pt]
    0 & 0 & 2\end{bmatrix*}
\end{align*}
where the error dynamics defined as $\Gamma$ is then resulted in the following,
\begin{align*}
    \Gamma = \begin{bmatrix}
    0 & 0 & -0.375 - l_1\\
    1 & 0 & -1.750 - l_2\\
    0 & 1 & -2.500 - l_3\end{bmatrix}
\end{align*}
so that the characteristic polynomial of the error $\Gamma$ is matched to desired eigenvalues of $\lambda_i$. These desired poles are $\lambda_i, 1\to 3$, comprising $-5, -8$, and $-10$, such that
\begin{align*}
    \det(sI - \Gamma) = \det\begin{bmatrix*}[r]
     s &  0 & 0.375 + l_1\\
    -1 &  s & 1.750 + l_2\\
     0 & -1 & s + 2.500 + l_3\end{bmatrix*}
\end{align*}
yielding,
\begin{align*}
    s^3 + (2.5 + l_3)s^2 + (1.75 + l_2)s + (0.375 + l_1)
\end{align*}
Those desired poles create the polynomial $\alpha_d = s^3 + 23s^2 + 170s + 400$ which is then compared to the above result. The following is to obtain the $\Psi_o$ (\ref{Eq 28}) leading to the observer gain $\Psi$ as denoted in (\ref{Eq 29}), therefore
\begin{align*}
    \Psi_o = \begin{bmatrix*}[r]
    399.625  \\
    168.250  \\
     20.500  \end{bmatrix*}, \quad \Psi = \begin{bmatrix*}[c]
    855 \\
    169.667 \\
    41 \end{bmatrix*}
\end{align*}
Finally, the dynamic of the system could be written as $\Gamma_d$ using the information of the matrices of $A, \Psi$, and $C$ as given,
\begin{align*}
    \Gamma_d = \begin{bmatrix*}[r]
    -0.5 &  0.0 & -855.000\\
     0.5 & -1.5 & -169.667\\
     0.0 &  1.5 &  -41.500\end{bmatrix*}
\end{align*}
and moving to the output residual, since it is assumed that $\frac{1}{4}<x_i(0)\leq 4$ with $i = 1\to 3$, the initial value being allowed in the estimation is $\hat{x}_i(0) = \frac{1}{4}$. From this, it can be concluded that the largest possible (upper-bound) value of the initial error $\hat{e}$ throughout the states is
\begin{align*}
    \hat{e} = 4 - \frac{1}{4} = \frac{15}{4}
\end{align*}
with the information in (\ref{Eq 31}), the upper-bound residual of the output is given below, such that
\begin{align}
    \bar{\varepsilon}(t) = \frac{78}{64}e^{-5t} - \frac{765}{64}e^{-8t} + \frac{807}{64}e^{-10t}
\end{align}
Since this upper-bound shows the actual time of the residual with the condition of $x_i(0) = 4$, the detection threshold $\varepsilon(t)$ should be design larger than that actual function $\bar{\varepsilon}(t)$ so as to maintain the failure detection. Thus, the larger upper-bound of $\bar{\varepsilon}$(t) is
\begin{align}
    \bar{\varepsilon}(t) = \frac{78}{64}e^{-5t} + \frac{765}{64}e^{-8t} + \frac{807}{64}e^{-10t}
\end{align}
which is able to set as the detection threshold. 

\section{simulation and Findings}
\begin{figure*}[t!]
	\centering
	\begin{subfigure}[t]{0.32\linewidth}
		\centering
		\includegraphics[width=\textwidth]{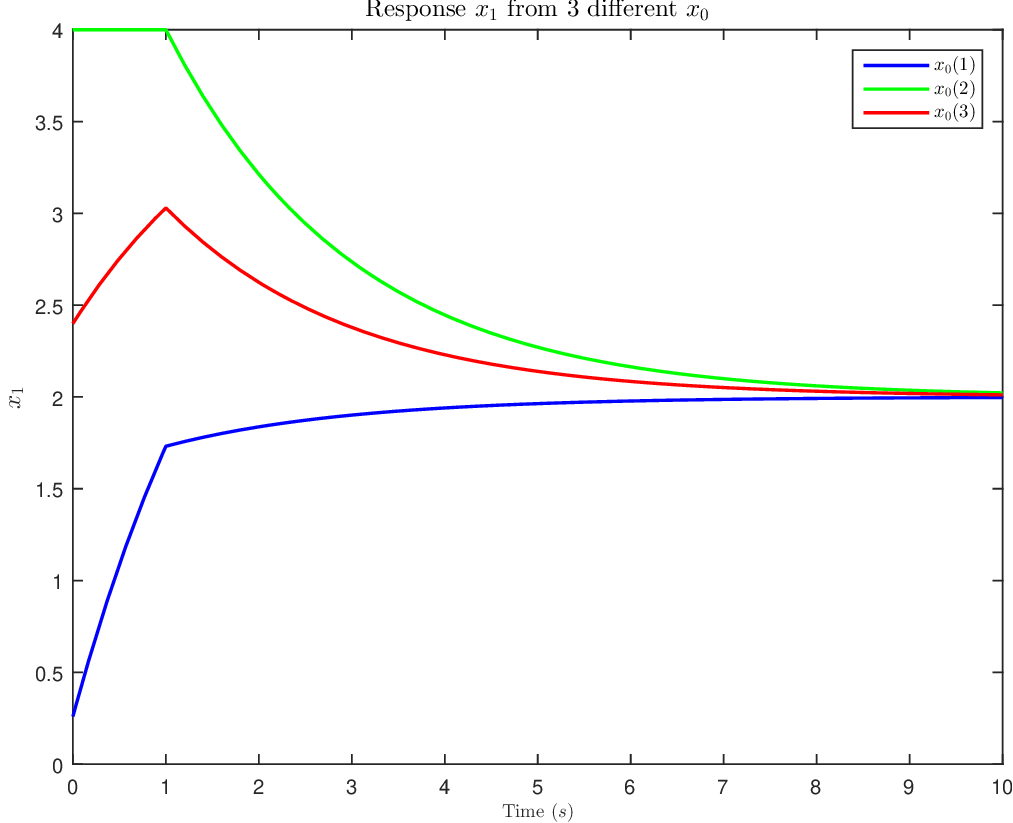}
		\caption{}
		\label{Fig 2a}
	\end{subfigure}
	~
    \begin{subfigure}[t]{0.32\linewidth}
		\centering
		\includegraphics[width=\textwidth]{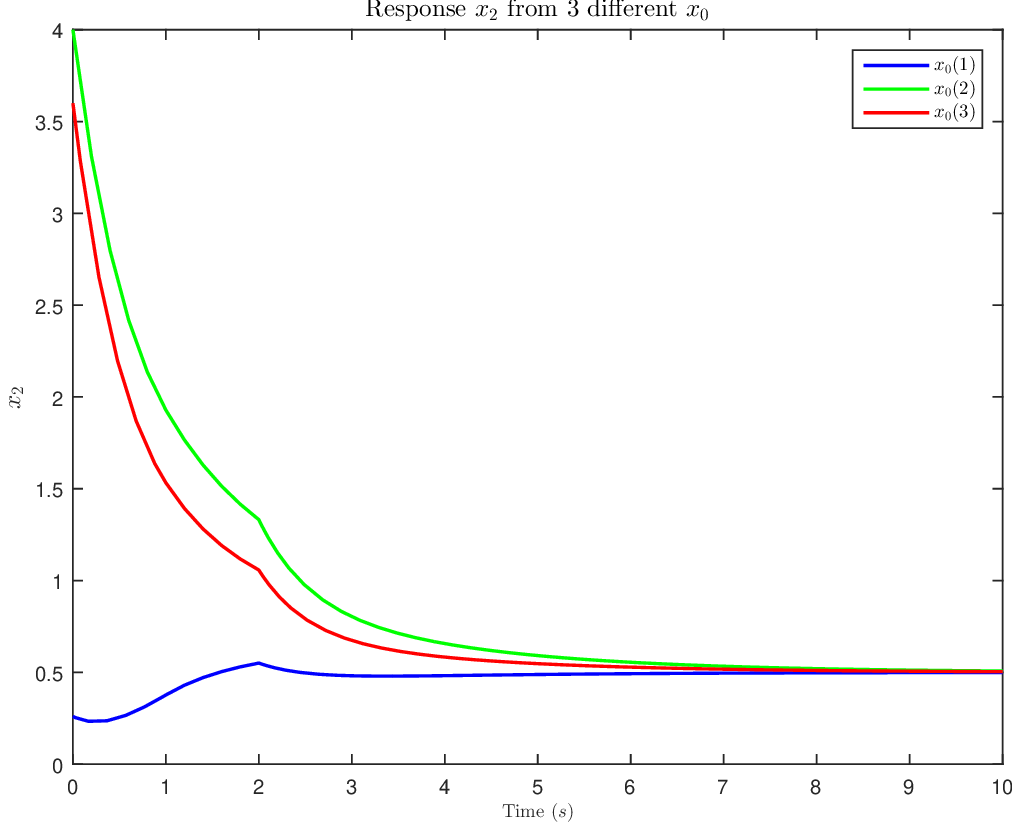}
		\caption{}
		\label{Fig 2b}
	\end{subfigure}
	~
    \begin{subfigure}[t]{0.32\linewidth}
		\centering
		\includegraphics[width=\textwidth]{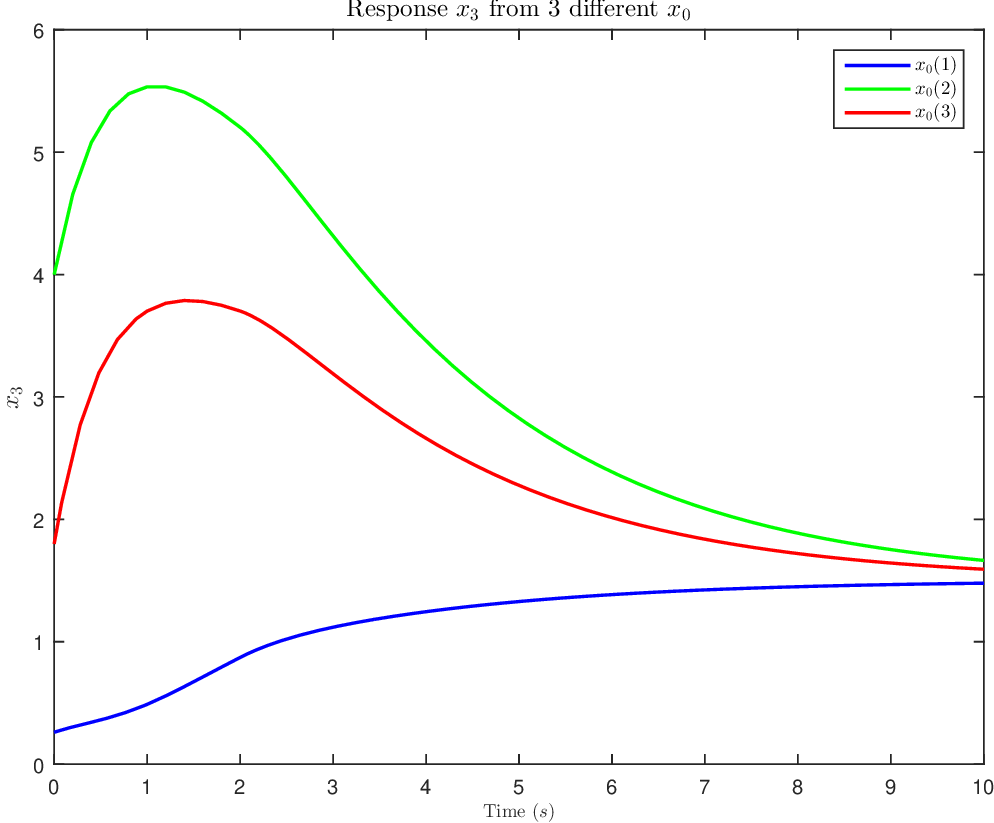}
		\caption{}
		\label{Fig 2c}
	\end{subfigure}\\
	\begin{subfigure}[t]{0.32\linewidth}
		\centering
		\includegraphics[width=\textwidth]{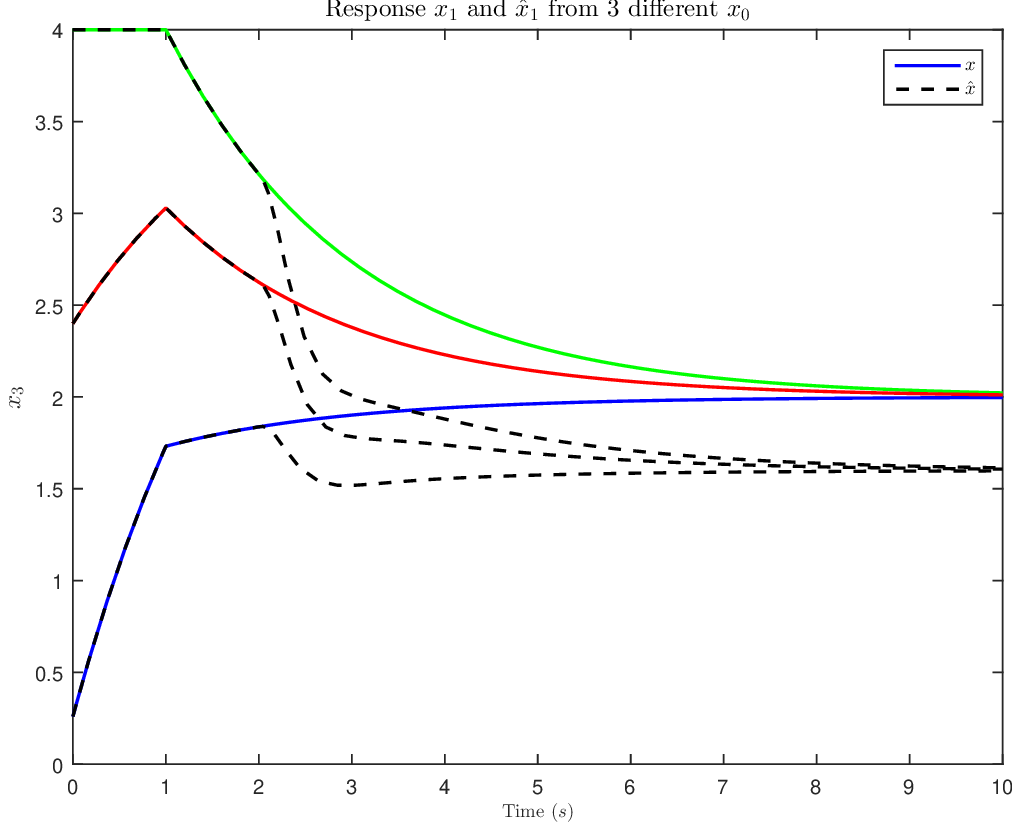}
		\caption{}
		\label{Fig 3a}
	\end{subfigure}
	~
    \begin{subfigure}[t]{0.32\linewidth}
		\centering
		\includegraphics[width=\textwidth]{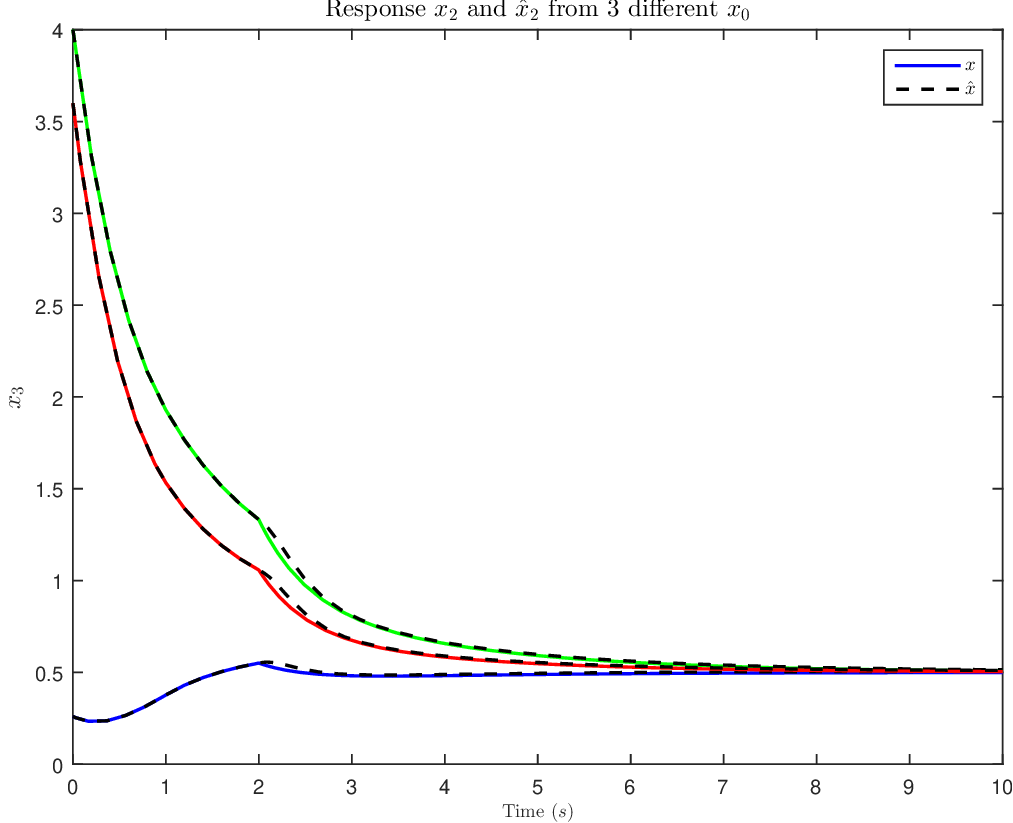}
		\caption{}
		\label{Fig 3b}
	\end{subfigure}
	~
    \begin{subfigure}[t]{0.32\linewidth}
		\centering
		\includegraphics[width=\textwidth]{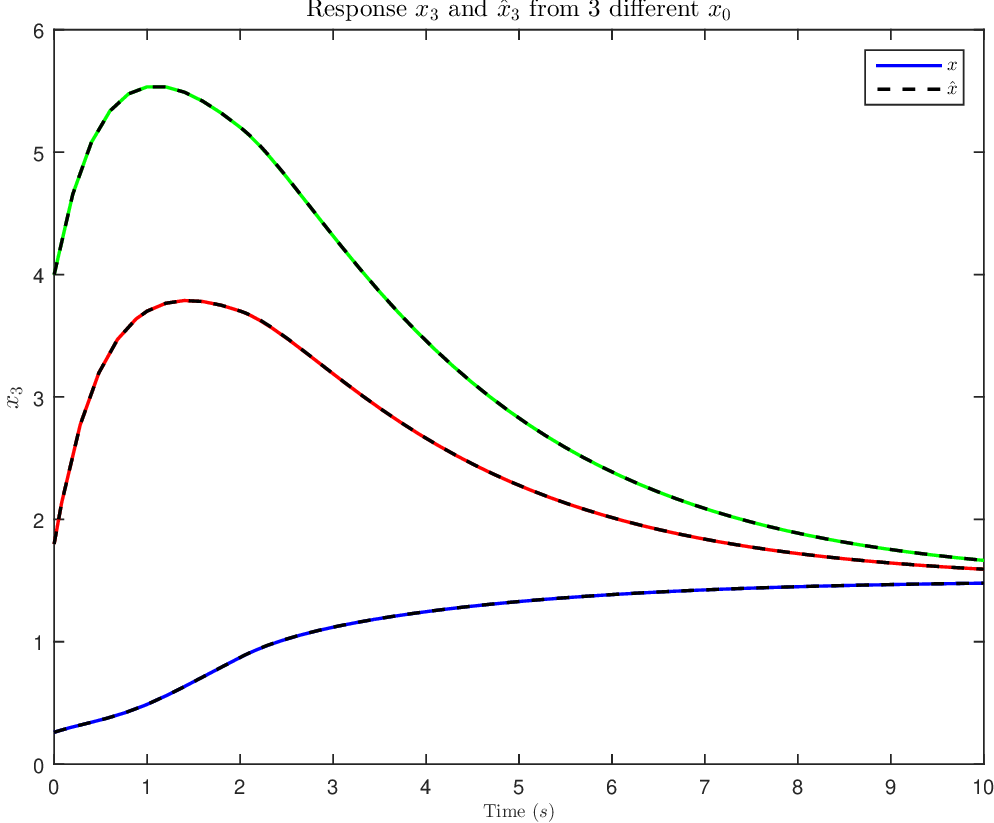}
		\caption{}
		\label{Fig 3c}
	\end{subfigure}\\
	\begin{subfigure}[t]{0.32\linewidth}
		\centering
		\includegraphics[width=\textwidth]{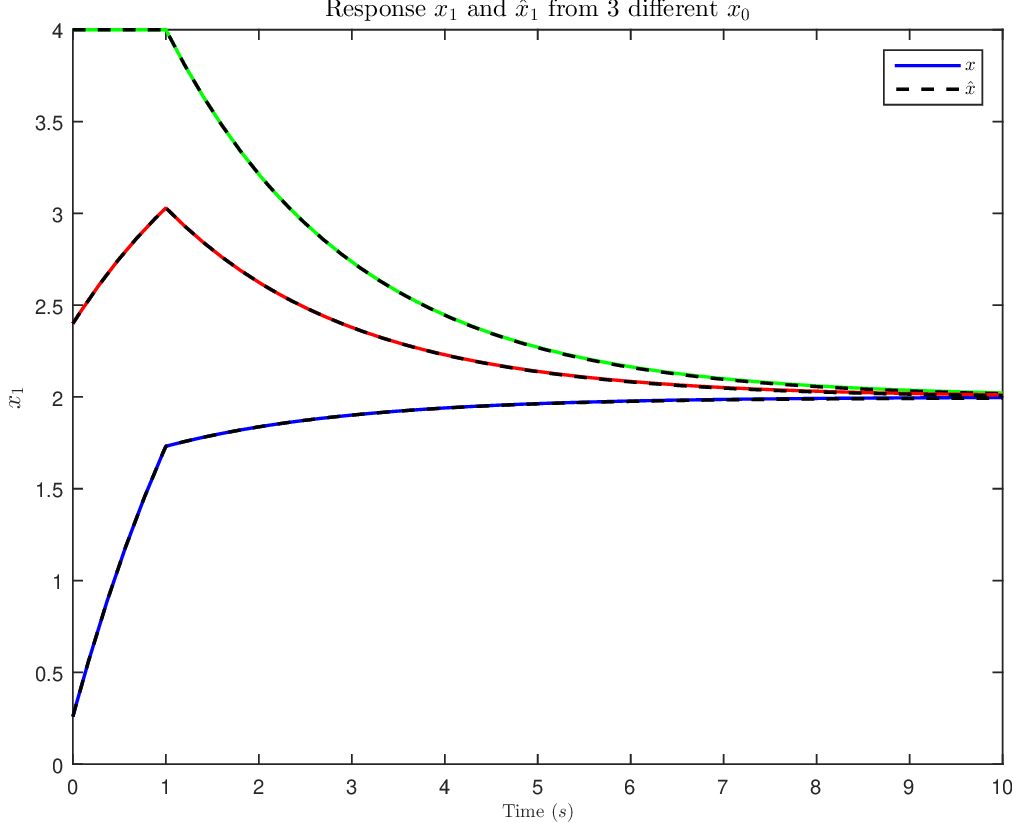}
		\caption{}
		\label{Fig 4a}
	\end{subfigure}
	~
    \begin{subfigure}[t]{0.32\linewidth}
		\centering
		\includegraphics[width=\textwidth]{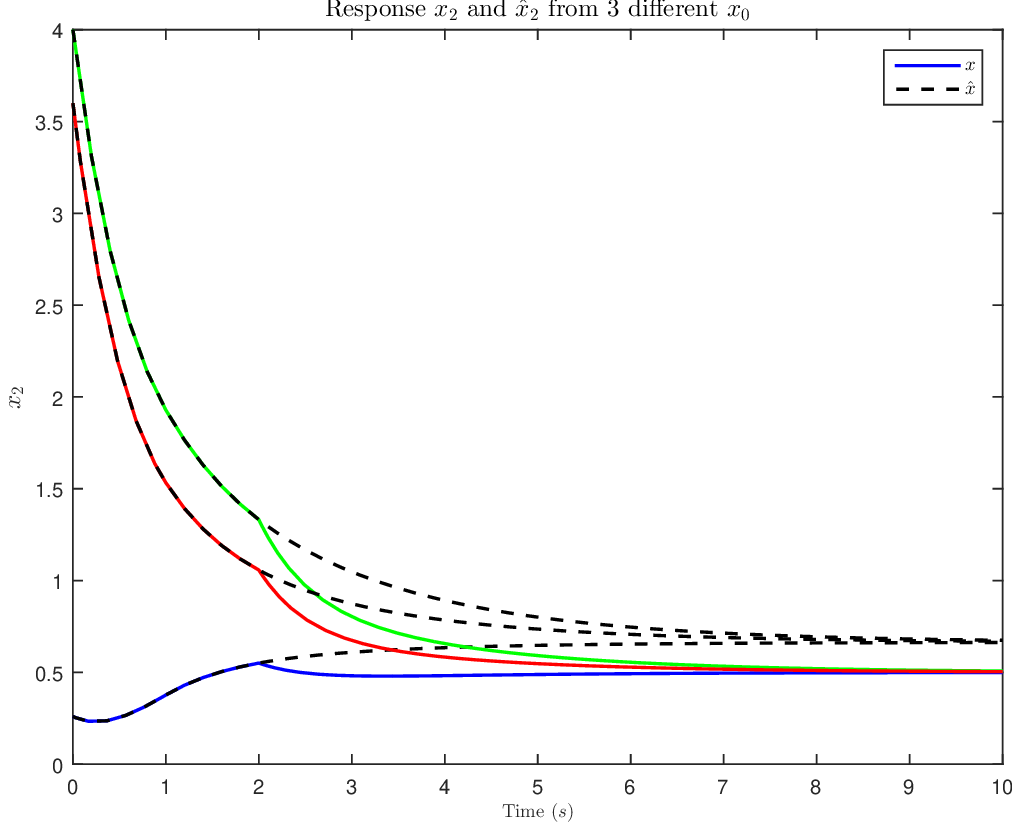}
		\caption{}
		\label{Fig 4b}
	\end{subfigure}
	~
    \begin{subfigure}[t]{0.32\linewidth}
		\centering
		\includegraphics[width=\textwidth]{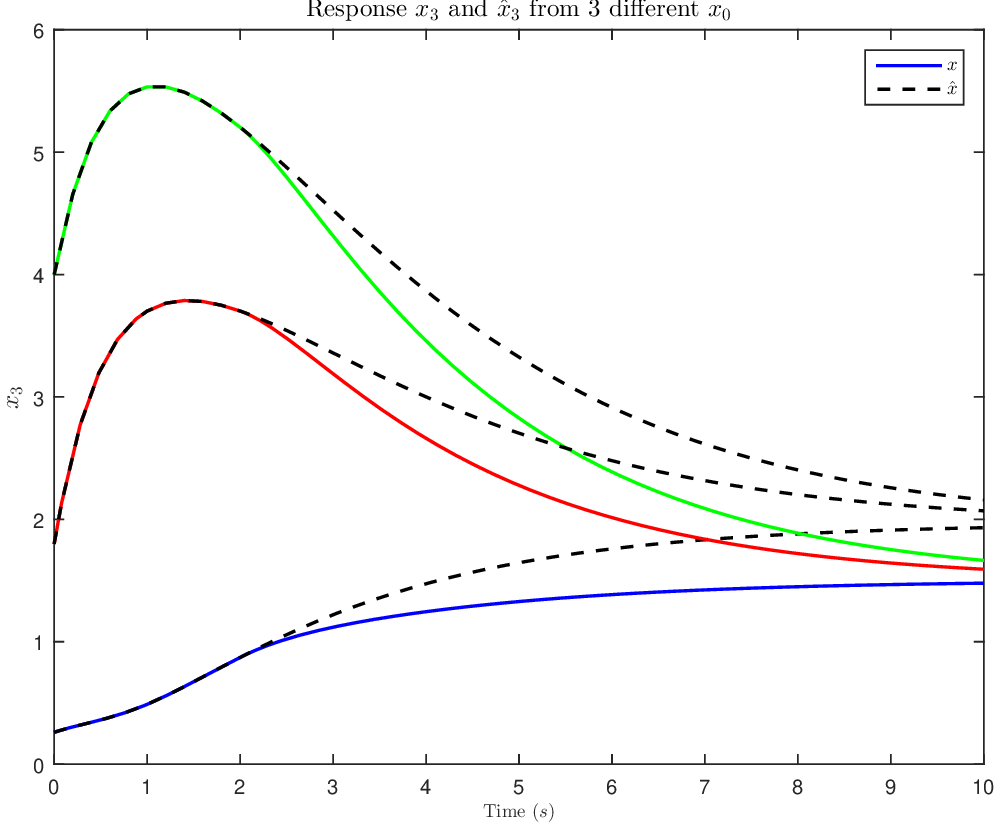}
		\caption{}
		\label{Fig 4c}
	\end{subfigure}\\
	\begin{subfigure}[t]{0.32\linewidth}
		\centering
		\includegraphics[width=\textwidth]{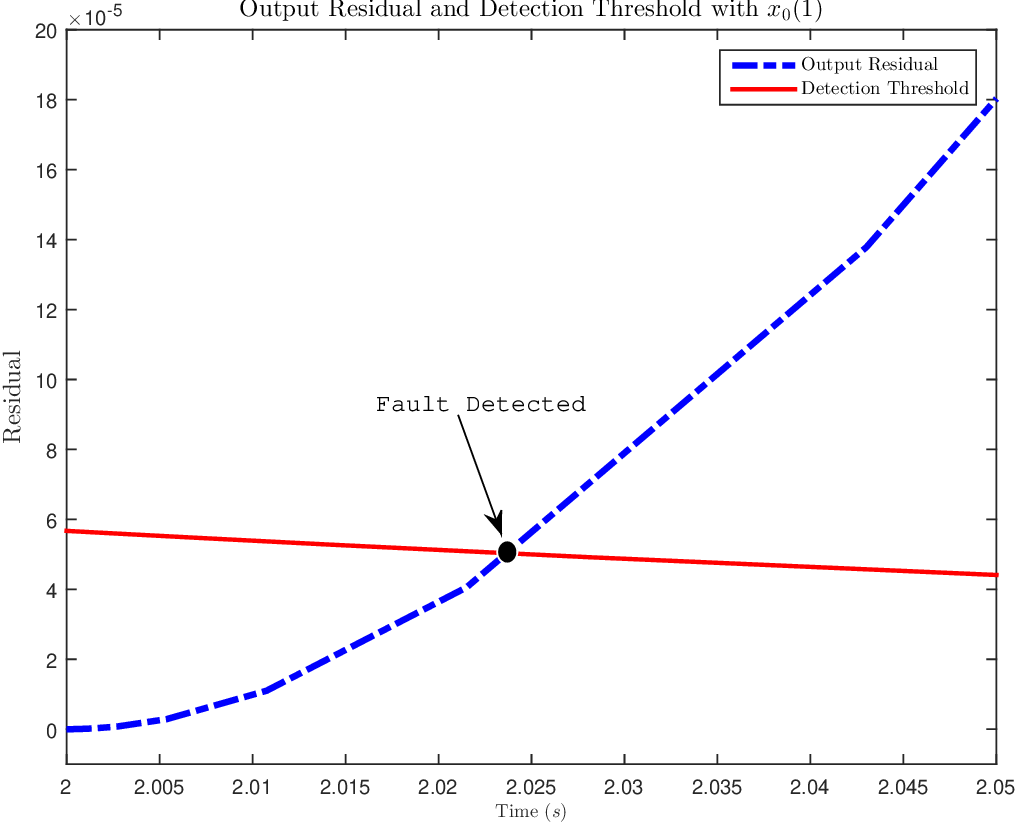}
		\caption{}
		\label{Fig 5a}
	\end{subfigure}
	~
    \begin{subfigure}[t]{0.32\linewidth}
		\centering
		\includegraphics[width=\textwidth]{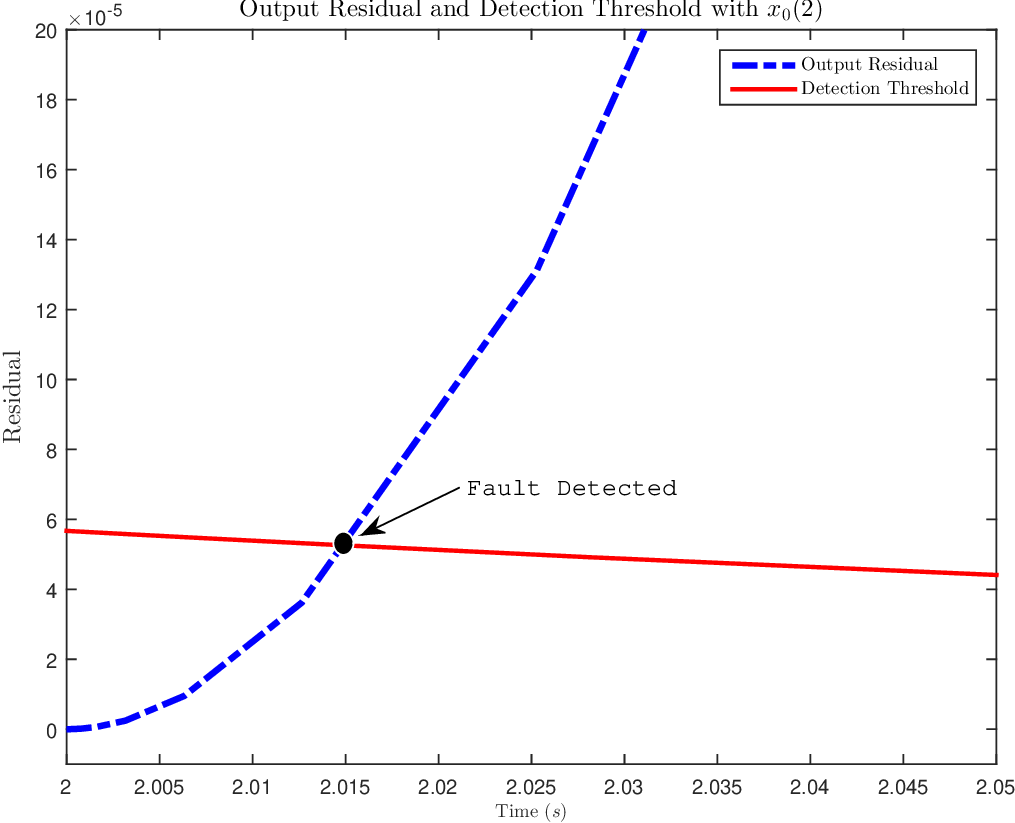}
		\caption{}
		\label{Fig 5b}
	\end{subfigure}
	~
    \begin{subfigure}[t]{0.32\linewidth}
		\centering
		\includegraphics[width=\textwidth]{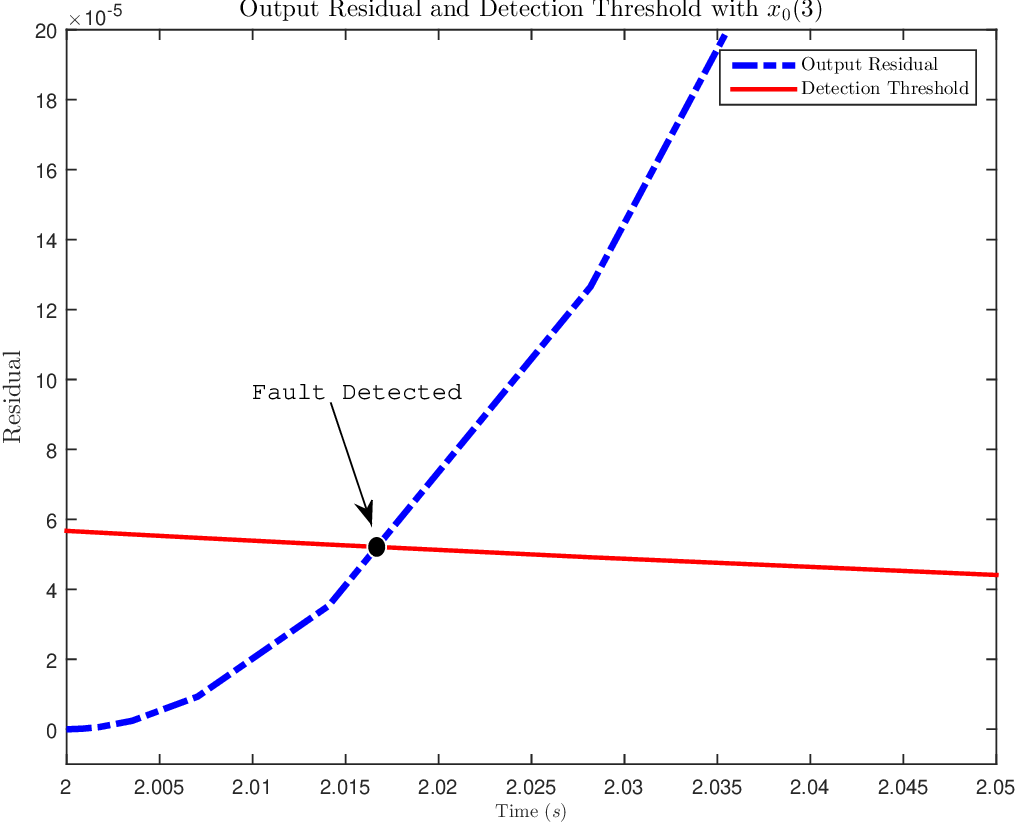}
		\caption{}
		\label{Fig 5c}
	\end{subfigure}
	\caption{Fig. \ref{Fig 2a}-\ref{Fig 2c} Response variations from $x_1$, $x_2$, and $x_3$ due to three initial conditions $x_0(1)$, $x_0(2)$, and $x_0(2)$; Fig. \ref{Fig 3a}-\ref{Fig 3c} The same scenarios with Luenberger observer; Fig. \ref{Fig 4a}-\ref{Fig 4c} The same scenarios with Adaptive-Scaling Kalman; Fig. \ref{Fig 5a}-\ref{Fig 5c} Fault detection from three divergent $x_0$} 
\end{figure*}

The following is to present the simulation results certain input $u(t)$ and $f(t)$ as in (\ref{Eq. 3}), such that,
\begin{align*}
    u(t) = \begin{cases}
    2, & 0\leq t \leq 1\\
    \\
    1, & t \geq 1
    \end{cases}
\end{align*}
with three different initial conditions $x_0$ which is used as a comparison apart from the estimation methods,
\begin{align*}
    \begin{bmatrix}
    x_0(1)\\
    x_0(2)\\
    x_0(3)
    \end{bmatrix} = \begin{bmatrix}
    0.26 & 0.26 & 0.26 \\
    4 & 4 & 4\\
    2.4 & 3.6 & 1.8
    \end{bmatrix}
\end{align*}
The design of gain observer along with two different estimation algorithms is compared in order to examine the ability to track with respect to MSE. Finally, the threshold design $\bar{\varepsilon}$ and residual output $\varepsilon$ is implemented to show where the fault is detected, questioning the $\bar{\delta}$ at $t_f \geq 2$. Fig. (\ref{Fig 2a}) presents the time behaviour of the state $x_1$ being affected by some initial conditions $x_0$ along with $u(t)$. Since there is no effect on the fault $f$, the response converges to the point of steady state. Due to the fault $f$ disturbance on the second tank $x_2$, the values plummet as soon as the fault is occurring at $t_f = 2s$. The changes in input $u(t)$ in $x_2$ seems no impact compared to that of $x_1$. $x_3$ which is free from the continuity of the fault also converges to the same point as $x_1$, showing the positive response of $u(t)$. 

It can be deduced that $x_0$ yields divergent transient of $\hat{x}$ because of the error steady state of $e(0) = x(0) - \hat{x}(0)$ with respect to certain initial conditions. Nevertheless, the steady states throughout $x$ cannot be said as the influence of $x_0$. Fig. (\ref{Fig 3a}), (\ref{Fig 3b}), (\ref{Fig 3c}) performs three different responses due to three various $x_0$ with disturbance observer. Compared to them, Fig. (\ref{Fig 4a}), (\ref{Fig 4b}), (\ref{Fig 4c}) implement adaptive-Kalman showing slightly divergent outcome which are better than that of consensus filtering. Fig. (\ref{Fig 5a}), (\ref{Fig 5b}), and (\ref{Fig 5c}) define the different fault detection where the output residual starts to overtake the threshold, meaning the fault is detected. From three $x_0(1), x_0(2), x_0(3)$, it results in $2.0235, 2.015$, and $2.0165$ respectively as the intersection between those two parameters, writing as $|\varepsilon(t_d)|>\bar{\varepsilon}(t_d)$. This can be summarized that the output residual would be relatively small for $t<t_f$ in comparison with the threshold. Due to the coincidence points after the transient observer, the time required to intersect the threshold is just slightly after the occurrence of the fault $f$.

\section{Conclusion}
The initial mathematical model of hydraulic system has been proposed along with the construction of stability properties. The estimation methods are applied with the Luenberger observer, adaptive-scaling Kalman, and consensus filtering to examine the unhealthy system. Fault detection scheme for certain time $t_f$ is designed and tested mathematically and in simulation. The results show that the fault detection scenarios are successfully proposed.

\section*{Acknowledgment}
Thanks to Professor Thomas Parisini from the Imperial College London who has taught me in the lecture leading to finishing this paper and to LPDP (Indonesia Endowment Fund for Education) Scholarship from Indonesia.

\bibliographystyle{IEEEtran}
\bibliography{reference.bib}

\begin{thebibliography}{10}
\providecommand{\url}[1]{#1}
\csname url@samestyle\endcsname
\providecommand{\newblock}{\relax}
\providecommand{\bibinfo}[2]{#2}
\providecommand{\BIBentrySTDinterwordspacing}{\spaceskip=0pt\relax}
\providecommand{\BIBentryALTinterwordstretchfactor}{4}
\providecommand{\BIBentryALTinterwordspacing}{\spaceskip=\fontdimen2\font plus
\BIBentryALTinterwordstretchfactor\fontdimen3\font minus
  \fontdimen4\font\relax}
\providecommand{\BIBforeignlanguage}[2]{{%
\expandafter\ifx\csname l@#1\endcsname\relax
\typeout{** WARNING: IEEEtran.bst: No hyphenation pattern has been}%
\typeout{** loaded for the language `#1'. Using the pattern for}%
\typeout{** the default language instead.}%
\else
\language=\csname l@#1\endcsname
\fi
#2}}
\providecommand{\BIBdecl}{\relax}
\BIBdecl

\bibitem{R1}
{Dingli Yu}, D.~N. {Shields}, and J.~L. {Mahtani}, ``A nonlinear fault
  detection method for a hydraulic system,'' in \emph{1994 International
  Conference on Control - Control '94.}, vol.~2, March 1994, pp. 1318--1322
  vol.2.

\bibitem{R2}
P.~{Bisták}, ``Identification and control of hydraulic system using visual
  feedback,'' in \emph{2016 International Conference on Emerging eLearning
  Technologies and Applications (ICETA)}, Nov 2016, pp. 29--34.

\bibitem{R3}
N.~Helwig, S.~Klein, and A.~Schütze, ``Identification and quantification of
  hydraulic system faults based on multivariate statistics using spectral
  vibration features,'' \emph{Procedia Engineering}, vol. 120, pp. 1225 --
  1228, 2015, eurosensors 2015.

\bibitem{R4}
S.~{Vásquez}, M.~{Kinnaert}, and R.~{Pintelon}, ``Active fault diagnosis on a
  hydraulic pitch system based on frequency-domain identification,'' \emph{IEEE
  Transactions on Control Systems Technology}, vol.~27, no.~2, pp. 663--678,
  March 2019.

\bibitem{R5}
F.~{Wang} and Y.~{Chen}, ``Dynamic characteristics of pressure compensator in
  underwater hydraulic system,'' \emph{IEEE/ASME Transactions on Mechatronics},
  vol.~19, no.~2, pp. 777--787, April 2014.

\bibitem{R6}
S.~Gayaka and B.~Yao, ``Fault detection, identification and accommodation for
  an electro-hydraulic system: An adaptive robust approach,'' \emph{IFAC
  Proceedings Volumes}, vol.~41, no.~2, pp. 13\,815 -- 13\,820, 2008, 17th IFAC
  World Congress.

\bibitem{R7}
S.~Sharifi, A.~Tivay, S.~M. Rezaei, M.~Zareinejad, and B.~Mollaei-Dariani,
  ``Leakage fault detection in electro-hydraulic servo systems using a
  nonlinear representation learning approach,'' \emph{ISA Transactions},
  vol.~73, pp. 154 -- 164, 2018.

\bibitem{R8}
G.~Shen, Z.~Zhu, J.~Zhao, W.~Zhu, Y.~Tang, and X.~Li, ``Real-time tracking
  control of electro-hydraulic force servo systems using offline feedback
  control and adaptive control,'' \emph{ISA Transactions}, vol.~67, pp. 356 --
  370, 2017.

\bibitem{R9}
Z.~MA, S.~WANG, J.~SHI, T.~LI, and X.~WANG, ``Fault diagnosis of an intelligent
  hydraulic pump based on a nonlinear unknown input observer,'' \emph{Chinese
  Journal of Aeronautics}, vol.~31, no.~2, pp. 385 -- 394, 2018.

\bibitem{R10}
K.~{Vereide}, B.~{Svingen}, T.~K. {Nielsen}, and L.~{Lia}, ``The effect of
  surge tank throttling on governor stability, power control, and hydraulic
  transients in hydropower plants,'' \emph{IEEE Transactions on Energy
  Conversion}, vol.~32, no.~1, pp. 91--98, March 2017.

\bibitem{R11}
D.~{Angeli}, ``A lyapunov approach to incremental stability properties,''
  \emph{IEEE Transactions on Automatic Control}, vol.~47, no.~3, pp. 410--421,
  March 2002.

\bibitem{R12}
D.~{Nesic} and A.~R. {Teel}, ``Input-output stability properties of networked
  control systems,'' \emph{IEEE Transactions on Automatic Control}, vol.~49,
  no.~10, pp. 1650--1667, Oct 2004.

\bibitem{R13}
Y.~{Huang}, Y.~{Zhang}, B.~{Xu}, Z.~{Wu}, and J.~A. {Chambers}, ``A new
  adaptive extended kalman filter for cooperative localization,'' \emph{IEEE
  Transactions on Aerospace and Electronic Systems}, vol.~54, no.~1, pp.
  353--368, Feb 2018.

\bibitem{R14}
V.~Lippiello, B.~Siciliano, and L.~Villani, ``Adaptive extended kalman
  filtering for visual motion estimation of 3d objects,'' \emph{Control
  Engineering Practice}, vol.~15, no.~1, pp. 123 -- 134, 2007.

\bibitem{R15}
S.~{Roujol}, B.~{Denis de Senneville}, S.~{Hey}, C.~{Moonen}, and M.~{Ries},
  ``Robust adaptive extended kalman filtering for real time mr-thermometry
  guided hifu interventions,'' \emph{IEEE Transactions on Medical Imaging},
  vol.~31, no.~3, pp. 533--542, March 2012.

\bibitem{R16}
M.~{Efe}, J.~A. {Bather}, and D.~P. {Atherton}, ``An adaptive kalman filter
  with sequential rescaling of process noise,'' in \emph{Proceedings of the
  1999 American Control Conference (Cat. No. 99CH36251)}, vol.~6, June 1999,
  pp. 3913--3917 vol.6.

\bibitem{R17}
V.~A. Bavdekar, R.~B. Gopaluni, and S.~L. Shah, ``Evaluation of adaptive
  extended kalman filter algorithms for state estimation in presence of
  model-plant mismatch,'' \emph{IFAC Proceedings Volumes}, vol.~46, no.~32, pp.
  184 -- 189, 2013, 10th IFAC International Symposium on Dynamics and Control
  of Process Systems.

\bibitem{R18}
R.~{Olfati-Saber}, ``Distributed kalman filter with embedded consensus
  filters,'' in \emph{Proceedings of the 44th IEEE Conference on Decision and
  Control}, Dec 2005, pp. 8179--8184.

\bibitem{R19}
M.~K. Wafi, ``Filtering module on satellite tracking,'' \emph{AIP Conference
  Proceedings}, vol. 2088, no.~1, p. 020045, 2019.

\end{thebibliography}

\end{document}